\documentclass[journal=nalefd,manuscript=letter]{achemso}

\usepackage{graphicx}
\usepackage{dcolumn}
\usepackage{bm}
\usepackage{graphicx}
\usepackage{subfig}
\usepackage{comment}
\usepackage{amsmath}
\usepackage{amssymb}
\usepackage[below]{placeins}
\usepackage[usenames, dvipsnames]{color}
\usepackage{siunitx}
\usepackage{float}
\usepackage{tabularx}
\usepackage{hyperref}
\usepackage{cleveref}
\usepackage{pdfpages}
\usepackage{lipsum}
\usepackage{soul}
\usepackage{xcolor}
\usepackage{comment}
\usepackage[font={small}]{caption}
\crefformat{figure}{#2{\color{blue}#1}#3}
\crefformat{equation}{#2{\color{blue}#1}#3}
\usepackage[mathscr]{eucal}
\DeclareMathAlphabet{\mathpzc}{OT1}{pzc}{m}{it}

\hypersetup{
colorlinks   = true, 
  urlcolor     = blue, 
  linkcolor    = black, 
  citecolor   = black 
}

\title{ Multistate Control of Nonlinear Photocurrents in Optoferroelectrics  via phase manipulation of light field
 }
\author{Ali Kazempour}
     \affiliation{Department of physics, Payame Noor University, PO BOX 119395-3697, Tehran, Iran}
      \affiliation{Department of Physics, Ulsan National Institute of Science and Technology, Ulsan, 689-798, Korea}

    \author{Esmaeil Taghizadeh Sisakht}
   \affiliation{Department of Physics, Ulsan National Institute of Science and Technology, Ulsan, 689-798, Korea}

     \author{Mahmut Sait Okyay}
   \affiliation{Materials Science and Engineering Program, Department of Chemical and Environmental Engineering, University of California-Riverside,
Riverside, CA 92521, USA}

   \author{Xiao Jiang}
   \affiliation{Department of Physics, Ulsan National Institute of Science and Technology, Ulsan, 689-798, Korea}

\author{Shunsuke Sato}
   \affiliation{Center for Computational Sciences, University of Tsukuba,Tsukuba 305-8577 Japan}

\author{Noejung Park}
   \affiliation{Department of Physics, Ulsan National Institute of Science and Technology, Ulsan, 689-798, Korea}
 \email{  noejung@unist.ac.kr}

\begin{document}

\begin{abstract}
Ultrafast optical control of ferroelectricity based on short and intense light can be utilized to achieve accurate manipulations of  ferroelectric materials, which may pave a basis for future breakthrough in nonvolatile memories. Here, we demonstrate that phase manipulation of electric field in the strong field sub-cycle regime induces a nonlinear injection current, efficiently coupling with the topology of band structure and enabling dynamic reversal of both current and polarization. Our time-dependent first-principles calculations reveal that tuning the phase of linearly or circularly polarized light through time-varying chirp, or constant carrier envelop phases within sub-laser-cycle dynamics effectively breaks the time-reversal symmetry, allowing the  control over current and electronic polarization reversal over multi-ferroelectric states. Our time- and momentum-resolved transverse current analysis reveal the significance of  Berry curvature  higher order poles in the apparent association between  the odd (even) orders of Berry curvature multipoles to odd (even) pseudo-harmonics in driving polarization dynamics reversal. We suggest that these phase manipulations of short pulse waveform may lead to unprecedented accurate control of nonlinear photocurrents and polarization states, which facilitate the development of precise ultrafast opto-ferroelectric devices.
\end{abstract}
\section{\ INTRODUCTION}

Ferroelectric materials, characterized by their spontaneous electric polarization that can be reversed by a bias, have recently gained significant interest, particularly for potential technological applications\cite{bian2024developing,wang2021two,kim2018application,scott1989ferroelectric,qi2021review,gao2021two,barraza2021colloquium}. In recent years, the exploration of opto-ferroelectric properties has opened new avenues for manipulating electronic polarization at ultrafast timescales\cite{guo2021recent,yang2018light}. The interaction of these materials with ultrashort and ultrastrong laser pulses allows high-speed generation of non-equilibrium carriers, leading to polarization dynamics, polarization switching, and domain evolution.\cite{yang2018light,lian2019indirect}. Such a high-speed accurate control of ferroelectric polarization may give rise to new functionalities in the fields of non-volatile memory elements, sensors, or transducers\cite{scott1989ferroelectric,guo2013non}.

Among two dimensional (2D) ferroelectric structures, optical control of polarization switching\cite{guo2021recent,guan2020recent} is of particular interest, and recent studies have shown promising advancements in this area, especially in group IV monochalcogenides\cite{fei2016ferroelectricity}. In SnTe monolayers, researchers have demonstrated reversible optical control of in-plane ferroelectric polarization using linearly polarized light. The mechanism was shown to involve the selective excitation of optical phonon modes in the terahertz regime that couple to the ferroelectric order parameter\cite{zhou2021terahertz}. Theoretical studies combining group theory and first-principles calculations have shown that intrinsic in-plane ferroelectricity in materials like GeS monolayers can induce spontaneous valley polarization, leading to the electrical control of ferrovalley states. The unique electronic characteristics of group-IV monochalcogenides are thought to enable the development of electrically tunable polarizers, leading to the generation of linearly or even circularly polarized light\cite{shen2017electrically}. Moreover, the inherent non-centrosymmetric nature of ferroelectric materials makes them ideal platform for nonlinear optical studies such as second harmonic generation and shift currents\cite{rangel2017large,orenstein2021topology,wang2017giant,jin2024peculiar}. Recent studies\cite{kim2019prediction} have introduced the photon helicity as a tool to control photogalvanic charge and spin currents via a ferroelectricity-driven adjustment of Berry curvature dipole, eventually leading to polarization reversal. For example, Wang et al. revealed that ferroelastic switching flips nonlinear shift and circular photocurrents by $\pm90^\circ$, while ferroelectric switching induces a $180^\circ$ flip. Switching light polarization between left and right circular also flips circular photocurrent by  $180^\circ$\cite{wang2019ferroicity}.
  Among the monochalcogenide class, SnTe has attracted the most wide interest. Monolayer SnTe exhibits strong in-plane ferroelectricity \cite{chang2016discovery} and significant Berry curvature effects due to their non-centrosymmetric structure\cite{jin2024peculiar}, making them highly sensitive to external phase modulation of light. This sensitivity, combined with their direct band gap and robust spin-orbit coupling, positions SnTe as an ideal candidate for exploring the dynamic control of electronic states, offering potential advancements in ultrafast optoelectronic devices.

A particularly promising approach involves manipulating chirp parameters and the carrier-envelope phase ($\phi_{CEP}$) for pulse shaping, which advantageously allows precise controls over electronic states transition\cite{neyra2016high,astapenko2009excitation} together with efficient excitation of nonlinear phonons through optimized pulse shaping.\cite{itin2018efficient}. $\phi_{chirp}$ modulates timing and intensity, aiding polarization reversal, while $\phi_{CEP}$ affects the symmetry and amplitude of the induced polarization. It has been shown that \cite{higuchi2017light,zhang2023residual} the residual current in graphene can be reversed by changing the $\phi_{CEP}$, attributed to Landau-Zener-Stückelberg (LZS) interference. Additionally, the complex $\phi_{CEP}$ dependence of solid harmonics, influenced by tunneling and multiphoton ionization, underscores the nuanced control achievable through $\phi_{CEP}$ manipulation\cite{wang2019complex}. These insights advance ultrafast optoelectronics and tailored experimental designs. In this study, we investigate the interaction of phase-modulated short pulses with ferroelectric monolayer SnTe. Our findings demonstrate that manipulating the phase of light through techniques such as chirp and $\phi_{CEP}$ serves as a critical control parameter for ferroelectricity-driven nonlinear photocurrent switching. Both chirp and $\phi_{CEP}$ break time-reversal symmetry, enabling the residual current to reveal the characteristics as nonlinear injection currents. We discuss the second-order residual transverse current and its reversal, which scales with the dynamical polarization and modulates the Berry curvature and its  multipoles induced by chirped pulses.

 \section{Results}
\subsection{Geometry and electronic structure of monolayer SnTe}
\begin{figure}[t!]
    \centering
    \includegraphics[width=1\textwidth]{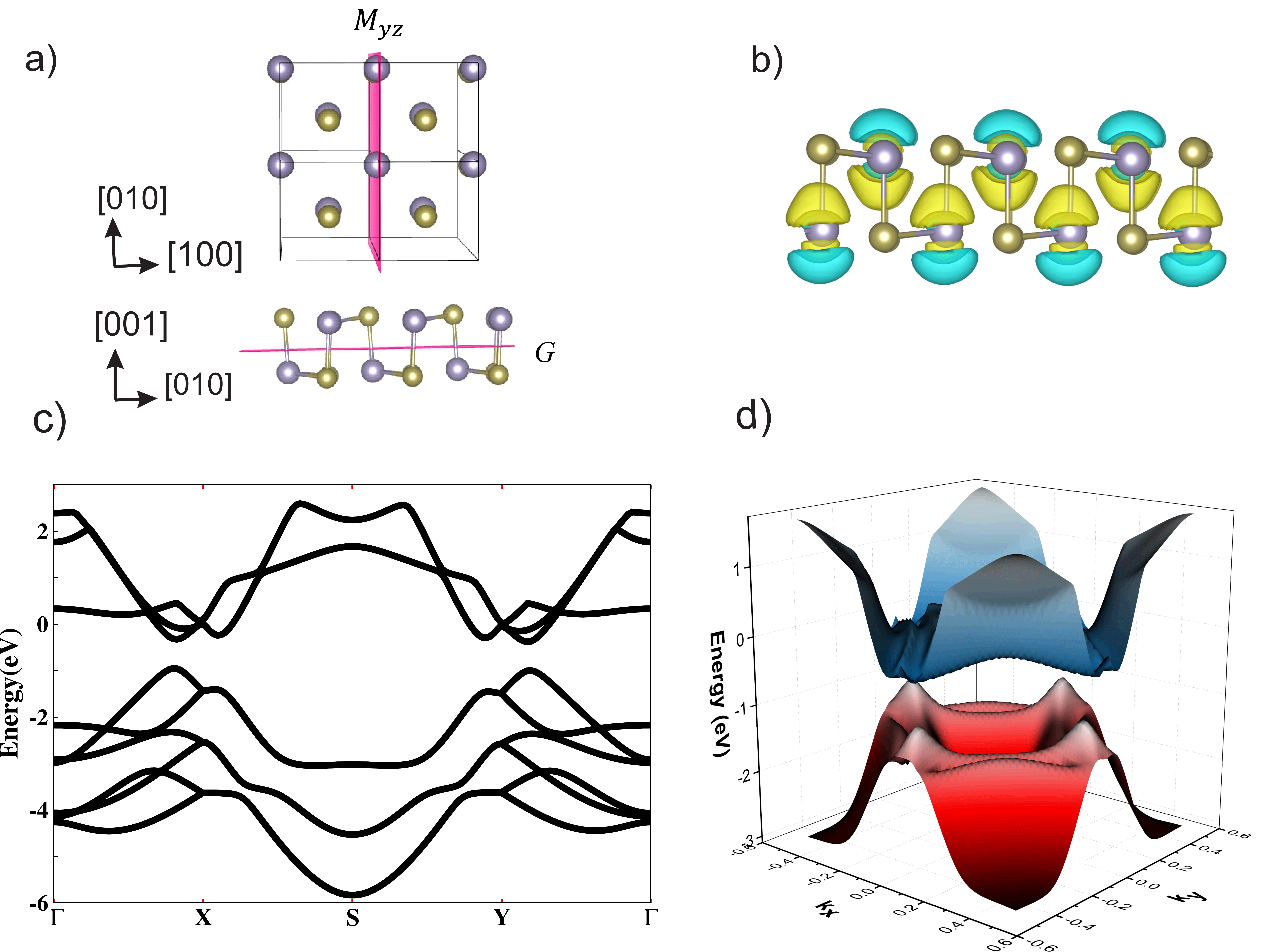}
    \caption{\textbf{Structural geometry, charge density, and electronic structure of the monolayer SnTe.}
    \textbf{(a)} Top and side views of the monolayer SnTe. Displacement along the y-axis between Sn (olive green) and Te (violet) induces in-plane ferroelectricity along the y direction.
    \textbf{(b)} Side view of the charge density difference between the ferroelectric phase and the reference paraelectric phase. Yellow/Cyan isosurfaces represent positive/negative charge distribution, with positive values showing increased electron density and negative values showing decreased electron density.
    \textbf{(c)} and \textbf{(d)} Two-dimensional and three-dimensional band structures, respectively, showing the inequivalent X and Y valleys.}
    \label{fig1}
\end{figure}

The monolayer SnTe features a staggered arrangement of tin (Sn) and tellurium (Te) atoms within a puckered rectangular lattice, characteristic of its structure, as depicted in Fig.~\cref{fig1}a. This monolayer belongs to the Pmn21 space group and exhibits in-plane ferroelectricity, making it structurally similar to the puckered form of phosphorene. In this configuration, Sn and Te atoms shift in opposite directions along the [010] axis (the y-axis in Fig.~\cref{fig1}a), results in broken inversion symmetry ($\mathcal{P}$) , thereby contributing to its unique electronic properties and spontaneous polarization. This structure exhibits glide $G$ and mirror $M_{yz} $ plane symmetries as shown in Fig.~\cref{fig1}a. These symmetries play a crucial role in shaping quantum geometry and nonlinear optical properties by imposing constrains on Berry curvature, enabling processes such as giant second harmonic generation\cite{fei2016ferroelectricity,wang2017giant,shin2020dynamical,kim2019prediction}. The ferroelectric polarization in the monolayer SnTe results in two distinct structural states, $P_y$ and $P_{-y}$, where the displacement of Sn and Te atoms along the [010] axis leads to polarization in opposite directions. The difference between these states lies in the direction of atomic displacement, which alters the electronic distribution and stabilizes one polarization state over the other depending on external conditions, such as strain or electric fields. When the ferroelectric polarization aligns with the armchair y-axis, the electronic structure of the monolayer SnTe reveals two valleys near the X and Y points in the Brillouin zone, henceforth referred to as the X and Y valleys. While these valleys exhibit similarities, they are not equivalent (Fig.~\cref{fig1}c,d). The monolayer SnTe behaves as a semiconductor with an indirect band gap of 0.7 eV (Fig.~\cref{fig1}c), as calculated using density functional theory (DFT). The valence band maximum and conduction band minimum occur at distinct points along the $\Gamma X$ and $\Gamma Y$ lines, respectively. The distinctive band structure around these valley points is responsible for the observed giant shift-current conductivities\cite{jin2024peculiar}.

\subsection{Multistate Control of Nonlinear Photocurrents by Phase Shaping }

We begin by calculating the evolution of current responses over time when exposed to a two-cycle short pulse with a constant carrier-envelope phase ($\phi_{CEP}$) and a time-varying $\phi_{chirp}$. The driver field is defined as $\bm {E}(t)=E_0 f(t) \cos(\omega t+\phi(t)) \bm{\hat{e}} $ where for chirped pulse $\phi(t)=\phi_{chirp}=bt^2$, while for a CEP-stabilized pulse, $\phi(t)=\phi_{CEP}$ is constant. Before making comparisons, we briefly explain the behavior of the pulse with $\phi_{chirp}$ and $\phi_{CEP}$. The $\phi_{CEP}$ which is the phase offset between the peak of the envelope and the phase of the carrier wave is time-independent. It allows the precise control of the electric field on a sub-cycle basis, which is crucial for the coherent control of electron dynamics in few-cycle pulses. Conversely, the $\phi_{chirp}$ refers to a time-dependent gradient in phase that controls the frequency content of the pulse over time, influencing the timing and spectral properties of the interaction. The symmetry of the pulse is partially disrupted by chirping, so the positive and negative parts are no longer identical. For instance, the acceleration caused by the interaction of an electron with the negative part of the oscillation is not canceled by the deceleration from the interaction with the corresponding positive part. As a result, there is a net acceleration, or equivalently, a net energy gain by the electron from the pulse.

\begin{figure}[t!]
    \centering
    \includegraphics[width=1\textwidth]{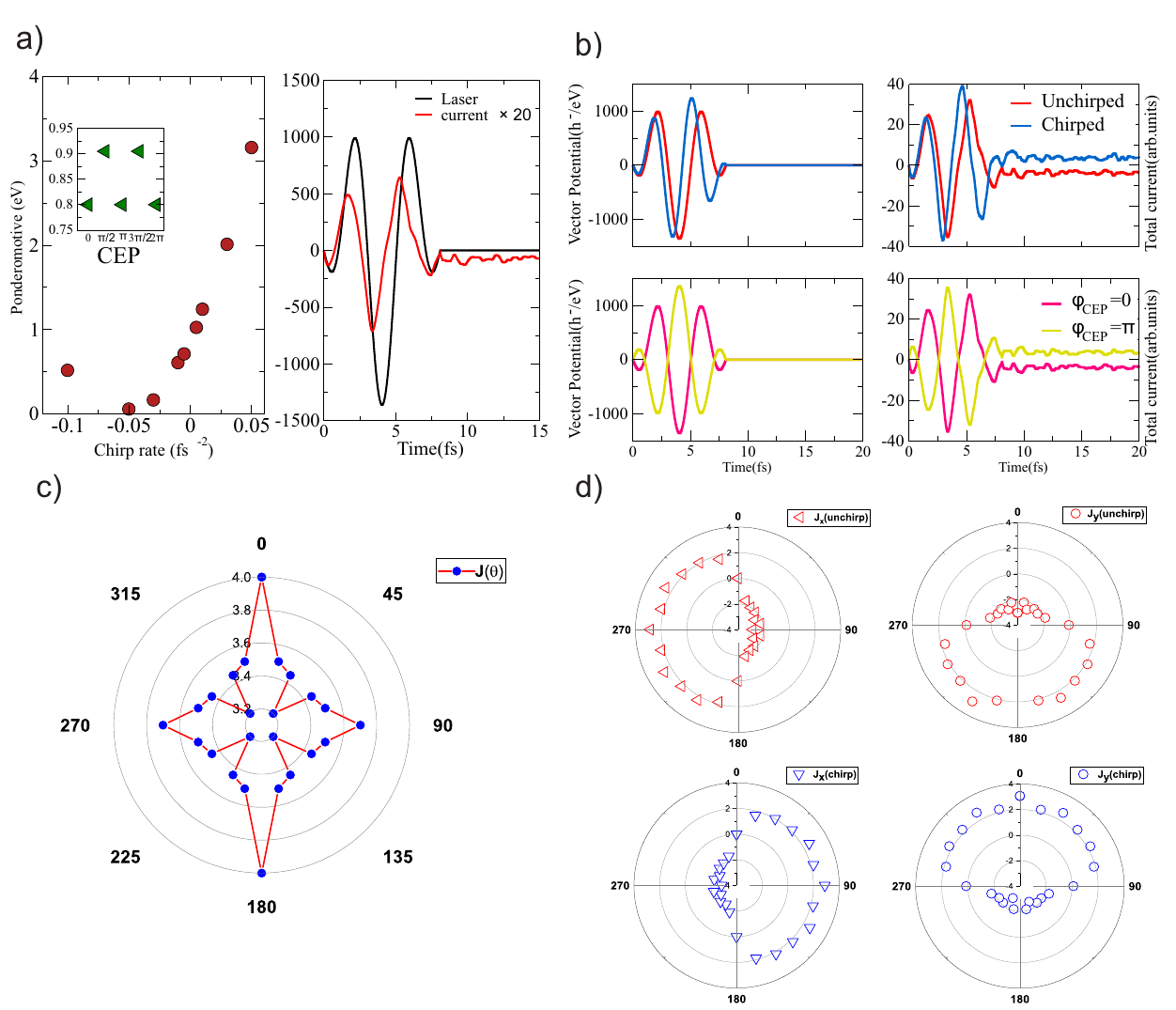}
    \caption{\textbf{ Residual current dynamics induced by chirped and CEP phase-modulated pulses }
    \textbf{(a)} Ponderomotive energy, defined as the average kinetic energy gained by electrons in a laser field, is shown as a function of constant $\phi_{CEP}$ and time-varying $\phi_{chirp}$. The changing frequency in the chirp case, modified the group velocity of electrons ( the velocity associated with wave packet dynamics in k-space), resulting in a non-uniform induced current. (left). The phase delay of the current(red) relative to the applied field (black) is depicted (right).
    \textbf{(b)} The top (bottom) panel shows the total current comparison between unchirped ($\phi_{CEP}$=0) and chirped ($\phi_{CEP}$=$\pi$) conditions. The residual current in the chirped case ($\phi_{CEP}$=$\pi$) is reversed compared to the unchirped case ($\phi_{CEP}$=0) when the laser ends.
    \textbf{(c)} Full polarimetry of the residual current magnitude for unchirped case..
    \textbf{(d)} The top (bottom) panel compares the transverse and longitudinal residual current components for unchirped (red) and chirped (blue) conditions, clearly showing the current direction reversal.
}
 \label{fig2}
\end{figure}

Figure.~\cref{fig2}a illustrates a typical comparison of Ponderomotive energy ($U_p=\alpha E^2/\omega^{2}$, where $\alpha$ is proportionality constant)  variation  over one cycle for an input frequency of 1.55 eV with $\phi_{chirp}$ and $\phi_{CEP}$ at an intensity of $10^{13} \, \text{W/cm}^2$. The relationship between the Ponderomotive energy and the Keldysh parameter $( \gamma )$, defined as $( \gamma = \sqrt{\frac{I_p}{2U_p}} , )$where $I_p$ is ionization potential, offers a quantitative method to differentiate multiphoton mechanism from the tunneling ionization mechanisms \cite{kruchinin2018colloquium,kazempour2021transient}. This relationship indicates that phase modulation can influence the relative contributions of these two ionization processes. Thus, a higher $U_p$ favors tunneling ionization, where intraband dynamics of electrons affect interband transitions\cite{wang2019complex,higuchi2017light}. Therefore, the relative magnitude of Ponderomotive energy gained by an electron under a linearly chirped laser alters both interband and intraband transition dynamics, making them much more complex compared to a $\phi_{CEP}$ modulated laser. Further, it should be noted that, while keeping the ionic energy fixed during the calculation, the energy of both configurations with equilibrium spontaneous polarizations $P_{+y}$ and $P_{-y}$ becomes unequal and lose balance after exposure to both unchirped and chirped irradiation. This destabilizes the double well potential surface, decreases the depth of the well, and leads to polarization reversal, which will be discussed later.

 Figure~\cref{fig2}b shows vector potential time-profile and the induced current response for linearly polarized pulses with
 unchirped/chirped and $\phi_{CEP}$ = 0, $\pi$ phases. The time-dependent Kohn-Sham equation for the electron dynamics and the definitions of the cell-averaged and momentum-resolved current are described in the Method section. We first align the electric field along the in-plane spontaneous polarization direction [010]  and the driving field is active only during the first two periods of oscillation and is turned off after the second period.
 The chirp phase is chosen  linear and  must change the sign of the vector potential before it reaches zero. As the driving fields for lasers in unchirped/chirped and $\phi_{CEP}$ = 0, $\pi$ phases are in the strong nonlinear regimes, the total currents resemble the laser shape with a small phase lag between the peaks of the current and the applied pulse. Increasing the incident frequency tends to reduce the phase delay. When the laser is turned off after the two-cycle operation, the current remains finite and almost constant as we neglect temperature and charge relaxation. due to remanent polarization, which is referred to as residual current in what follows. This dc residual current at sufficiently large field remains finite due to higher-order nonlinear process contribution\cite{mao2024nonlinear}.
Noticeably, the residual current induced by chirped light is opposite to unchirped light, as shown in Fig.~\cref{fig2}b (upper right panel). The reversal of residual current is also observed when the $\phi_{CEP}$ of the applied field changes from 0 to $\pi$, as shown in Fig.~\cref{fig2}b (bottom right panel).  In the Supplementary material we have shown that a sequential reversal of residual current occurs as a function of chirp rate for both positive and negative values, provided that the vector potential changes sign before being switched off. Breaking the time-reversal symmetry between Kramers pairs in the Brillouin zone using the chirp parameter induces a population imbalance. Referenced to the free chirped light, those positive chirp rates $b$ that generate a greater than an odd number but less than the next even number lead to a reversed current.

The full polarimetry profile of the total current, along with its longitudinal and transverse components for both unchirped and chirped lasers, is illustrated in Fig.~\cref{fig2}c,d. A transverse current ($J_{\perp}$) is generated alongside the longitudinal current ($J_{\parallel}$) at all incident angles, except for $\theta = 0$ and $\theta = \pi$, which correspond to the orientation of the spontaneous polarization.
  Both residual $J_{\perp}$ and $J_{\parallel}$ currents show a reversal after the laser pulse ends.  The symmetry of mirror $M_{yz}$ and also a two-fold rotation (C$_{2v}$) is clearly recovered from the polar plot of the total current. The $C_{2v}$ symmetry imposes constraints $\chi^{(2)}_{ijk} = -\chi^{(2)}_{jik}$  for $i \neq j$ on the nonlinear optical tensors such as second-order susceptibility $\chi^{(2)}$. The transverse component, which we show in the next section to be a second-order current, suggest that the spontaneous polarization, normalized to its static equilibrium value $P_s(\tau)/P_s(0)$, undergoes reversal in these transient regimes. More importantly, we will later address how ferroelectricity is dominantly affected by the electric field induced Berry curvature\cite{ye2023control,lai2021third}. Symmetry breaking or reduction, induced by chirp modulation of the phase, manifests in optical responses through the involvement of both odd and even order derivatives of the Berry curvature. 

Notably, selecting chirp rates \textit{b} of equal magnitude but opposite signs in applied electric field in the form of $\bm {E}(t)=E_0 f(t) \cos(\omega t+bt^2) \bm{\hat{e}} $ results in distinct residual currents, differing both in strength and direction, as illustrated in the Supplementary Material. This occurs because $\pm b$  produce different frequency spectra over the vector potential. We also present calculations for unchirped and chirped setups over a larger number of cycles, including the effects of pumped IR-active phonons in the structure, which amplify the dynamical polarizations. All calculations demonstrate a reversal in the direction of remanent currents, provided the system remains in the short-pulse regime, where both time-reversal $\mathcal{T}$ and inversion symmetries $\mathcal{P}$ are broken, though the magnitude of the residual current may vary. Furthermore, we observe that scattering mechanisms, such as phonons, do not significantly alter the overall outcome. For instance, we explored the enhancement of in-plane polarization via nonlinear phonon pumping, as described in \cite{shin2020dynamical}, and found that introducing out-of-plane $A_u$ and in-plane $E_u$ phonons does not impact the general trend. Notably, this residual current reversal is robust, persisting for up to 250 fs within the framework of Ehrenfest dynamics, even in the presence of ionic motion (see Supplementary Material Note 3).

In both unchirped and chirped pulses, the longitudinal ($J_{\parallel}$) and transverse ($J_{\perp}$) currents exhibit reversal. This sign change induces a $\pi$ polarization rotation, with similar current reversals expected for constant $\phi_{CEP}$ values of 0 and $\pi$. A parallel phenomenon occurs in Landau-Zener-Stückelberg interference in graphene\cite{higuchi2017light}, where the current direction depends on the $\phi_{CEP}$. The $\phi_{CEP}$ modulates electron trajectory lengths in reciprocal space, shifting interference from constructive to destructive due to population imbalances. This $\phi_{CEP}$ dependent interference explains the photocurrent's sensitivity to $\phi_{CEP}$ in graphene under few-cycle pulses.

\begin{figure}[t!]
    \centering
    \includegraphics[width=1\textwidth]{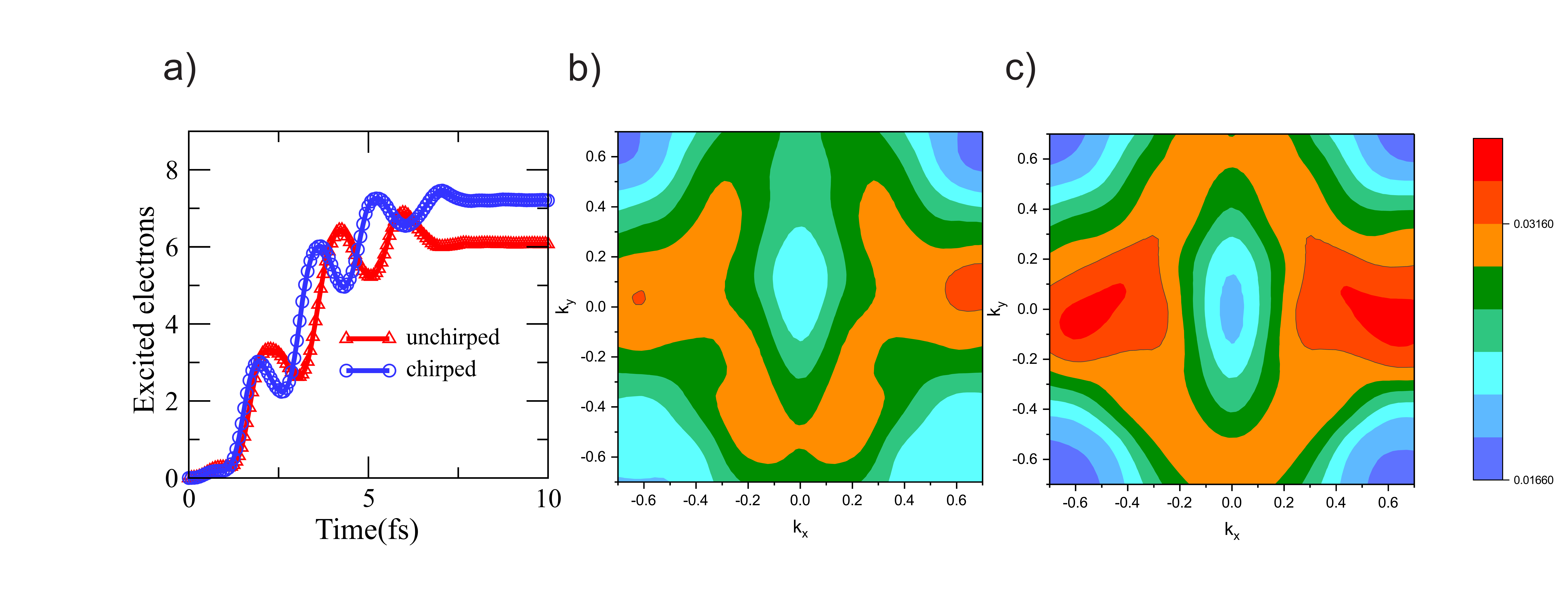}
   \caption{\textbf{Dynamics of excited electrons and population imbalance.}
    \textbf{(a)} The evolution of excited electron numbers in the conduction band over time for both unchirped (red) and chirped (blue) pulse conditions. The chirped pulse induces a higher population of excited electrons due to phase modulation effects, leading to an imbalance that enhances the residual current. \textbf{(b, c)} The distribution of $k$-resolved electron population within the first Brillouin zone (BZ) for unchirped  and chirped  fields, respectively. 
}
 \label{fig3}
\end{figure}

As shown in Fig.\cref{fig2}b, the time-reversal symmetry $\mathcal{T}$ of the vector potential is broken in the case of chirped pulses with a carrier-envelope phase ($\phi_{CEP}$) of $\pi$, where $\bm {A}(t) \neq \bm {A}(-t)$. This symmetry breaking in the vector potential leads to a time-reversal symmetry violation in the Hamiltonian, resulting in a population imbalance and a residual current.
Figure~\cref{fig3}a, b presents a comparison of the number of excited electrons over time and the resulting population imbalance across the Brillouin zone (BZ) for both unchirped and chirped pulses.  We attribute the observed population imbalance in the unchirped case to the intrinsic breaking of inversion symmetry within the system, which we trace back to the relaxation of accumulated polarization. This initial imbalance differs from the induced imbalance in the population during the aforementioned phases. Notably, the chirped pulse excites an additional electron into the conduction band compared to the unchirped case, leading to a distribution that initially favors the $-k_x$ valley at lower intensities but extends to other regions beyond the valleys in the strong-field regime. Additionally, as demonstrated in the Supplementary Information, chirp rates that generated number of excited electrons is between an odd and the next even integer reverse the direction of the current compared to that induced by an unchirped field. It is worth noting that the incident time of the interaction is short enough, allowing the system to remain below the damage threshold despite the presence of a large number of excited electrons.

While earlier studies have largely explored the connection between current direction, ferroelectricity, and the Berry curvature poles in monolayer SnTe~\cite{kim2019prediction,wang2019ferroicity}, our findings here provide a clearer perspective. We show that the reversibility of ferroelectric polarization—closely tied to current reversal through a sign change in Berry curvature—can be precisely controlled. This control is achieved by adjusting the population imbalance between the left and right valleys, which is governed by the handedness of circularly polarized light~\cite{kim2019prediction,wang2019ferroicity}.

\subsection{Chirp-driven time-reversal breaking and persistent injection current}

Sato et al.~\cite{sato2024limitations,mao2024nonlinear} investigated the limitations of mean-field methods, such as TD-Hartree, TD-Hartree-Fock, and TDDFT, in capturing second-order shift and injection currents. These mean-field approximations can result in unphysical DC responses, even after the system is exposed to linearly polarized light. This occurs due to population imbalances between $k$ and $-k$ states (Kramers pairs) in the Brillouin zone. Such imbalances are thought to arise quantum interference incurred on artificial self-excitation pathways, which are absent in exact many-body treatments. In TDDFT, such unphysical part of the residual current are attributed to time-nonlocal memory effects which can be mitigated by improving the exchange-correlation kernel $f_{Hxc}(r,r',\omega$).
Using a many-body Schrödinger framework, the study identified three key factors contributing to population imbalance: the vector potential, its interference with the induced mean-field potential, and the induced mean-field potential itself. Under time-reversal symmetry, these contributions cancel out; however, chirped laser pulses break this symmetry, leading to a real, physical injection current that persists after the pulse ends.

Furthermore, the introduction of a chirp imposes distinct phase shifts across pulse components, leading to quasi-phase matching that enhances quantum interference between excitation pathways \cite{kovacs2012quasi}. This mechanism can result in a more pronounced and stable residual current, in contrast to scenarios where time-reversal symmetry $\mathcal{T}$ remains preserved. We highlight that both the magnitude and direction of the induced dc current are highly sensitive to the specific properties of the chirped pulse, such as the chirp rate and the overall temporal profile.

Recently, the mechanisms underlying the generation of shift and injection currents under various material symmetries and light polarizations have been investigated\cite{xu2021pure}. It was demonstrated that in systems with broken $\mathcal{P}$, $\mathcal{T}$, and $\mathcal{PT}$ symmetries, both shift and injection currents can be produced simultaneously under LPL and CPL irradiation. Our findings confirm that introducing a chirp phase offers a framework that facilitates the concurrent generation of shift and injection currents. However, we identify that the residual current is an injection current, oriented orthogonally to the spontaneous polarization direction \cite{panday2019injection}. While the introduction of a chirp modifies the excitation of charge carriers across various energy levels, the injection current remains perpendicular to the polarization. This occurs because the chirped pulse affects the occupation of states within BZ without inducing strain or altering  electronic structure.

\subsection{Origin of Chirp-Induced Current and Polarization Switching }

The fundamental characteristics and switching dynamics of shift and injection currents in group-IV monochalcogenide monolayers are determined by the interplay between symmetries, shift vectors, and Berry curvature \cite{wang2019ferroicity,fregoso2017quantitative}. For instance, the ferroelectric properties, essential for modulating the Berry curvature and other nonlinear optical responses, are significantly influenced by the helicity of incident circular light and the resultant photogalvanic current \cite{kim2019prediction}. Moreover, in systems where the time-reversal symmetry $\mathcal{T}$ is preserved, the NLHE in metallic systems arises as a second-order response to external electric fields and is intimately connected with the Berry curvature dipole (BCD) of the Fermi surface \cite{sodemann2015quantum}. Furthermore,  the second-order harmonic generation resulting from transverse nonlinear optical responses in a $\mathcal{T}$-preserved insulator bears resemblance to the nonlinear Hall effect (NLHE) observed in metallic systems \cite{okyay2022second}. As the incident field intensity increases, multipoles of the Berry curvature become more pronounced. These terms, representing more intricate asymmetries in momentum space, grow increasingly important under stronger perturbations. When an external electric field excites additional electrons into the conduction band, the Fermi level rises, rendering the system more metallic in nature. We show that applying a chirped pulse offers an effective way to unveil the interplay of aforementioned factors by incorporating both odd and even orders of Berry curvature multipoles through the breaking of $\mathcal{T}$ symmetry.
  The obtained results can be explained in terms of interband Berry curvature, as described by \cite{okyay2022second}. In the presence of a phase-modulated strong field given by $\bm{E}(t) = E_0 \, \text{Re} [e^{i\Phi(t)}] \hat{\bm{x}}$, where $\Phi(t) = \omega t + \phi(t)$, we introduce the step function $\Theta(\Phi(t) - \omega_{nm})$ to account for all states that satisfy the gap frequency condition $\Phi(t) \geq \omega_{nm}$ where n and m are valence and conduction bands. In the strong-field regime, the anomalous velocities arising from both the valence and conduction bands are governed by the total Berry curvature of the system . This can be expressed in terms of the total interband Berry curvature $\Omega_{nm}(\bm{k})$ \cite{okyay2022second}. 


Using the corresponding vector potantial $\bm{A}(t)=\frac{-cE_0}{\Phi'(t)} Im[e^{i\Phi(t)}]\hat{\bm{x}}$, the phase-dependent Bloch oscillation of carriers is written as $\Delta {k}=\frac{-eE_0}{\hbar \Phi'(t)} Im[e^{i\Phi(t)}]$. This can be transferred to the interband Berry curvature and  consequently, the transverse current is given by:
\begin{equation}
  \mathbf{J}_{\perp}(t) = 2 \frac{e^2}{\hbar} \mathbf{E}(t) \times \sum_{nm} \int \frac{d^2 \mathbf{k}}{4\pi^2} \mathbf{\Omega}_{nm}(\mathbf{k} + \Delta k \hat{x}) \Theta(\Phi(t) - \omega_{mn}(\mathbf{k})).
\end{equation}

Depending on the form of phase $\Phi(t)$, the phase-modulated strong field can either preserve the  $\mathcal{T}$ symmetry or break it.  In former case, as we show in supplementary material, the transverse current $\mathbf{J}_{\perp}(t)$ is determined by the odd order derivatives of interband Berry curvature and as a result we revealed the interrelationship between even-order psudo-harmonics $\Phi(t)$ and series of even-order interband Berry curvature multipoles $\mathbf{D}^{(2p)}_{nm}(\Phi) = \int_{\mathbf{k}} \frac{\partial^{(2p-1)} \Omega_{nm}}{\partial \mathbf{k}^{(2p-1)}} \Theta(\Phi - \omega_{nm})
$ as following

\begin{equation}\label{even-sum}
\mathbf{J}_{\perp}(t) =\sum_{p=even,\geq 2}\sum_{q\geq p} \operatorname{Im}[e^{ip\Phi}]   \frac{ \alpha_q E^q_0}{\Phi'^{(q-1)}} D^{(q)}
\end{equation}

Here, we take $\alpha_q$ as constant coefficients and $\Phi'$ is time-derivative of phase. In addition, the above results show how different even-order nonlinear responses contribute to the transverse current. It is worth noting that if we consider the weak field limit in which the interband transition is dominant within effective two-band model, the above generalized equation is reduced to well-known second-order response of $\mathbf{J}_{\perp}(t)\propto   \widehat{z} \times {\bm{E}} ({\bm{D}_{nm}^{(2)}} \cdot {\bm{E}})$ \cite{sodemann2015quantum}. On the other hand, when $\mathcal{T}$ is broken the transverse current is given by (see supplementary material)

\begin{equation}\label{even-odd-sum}
\mathbf{J}_{\perp}(t) =\sum_{p=odd,\geq 1}\sum_{q \geq p}  \operatorname{Re}[e^{ip\Phi}] \frac{ \alpha_q E^q_0}{\Phi'^{q-1}} D^{(q)}+\sum_{p=even,\geq 2} \sum_{q \geq p} \operatorname{Im}[e^{ip\Phi}]   \frac{\alpha_q E^q_0}{\Phi'^{q-1}} D^{(q)}.
\end{equation}

\begin{figure}[t!]
    \centering
    \includegraphics[width=.95\textwidth]{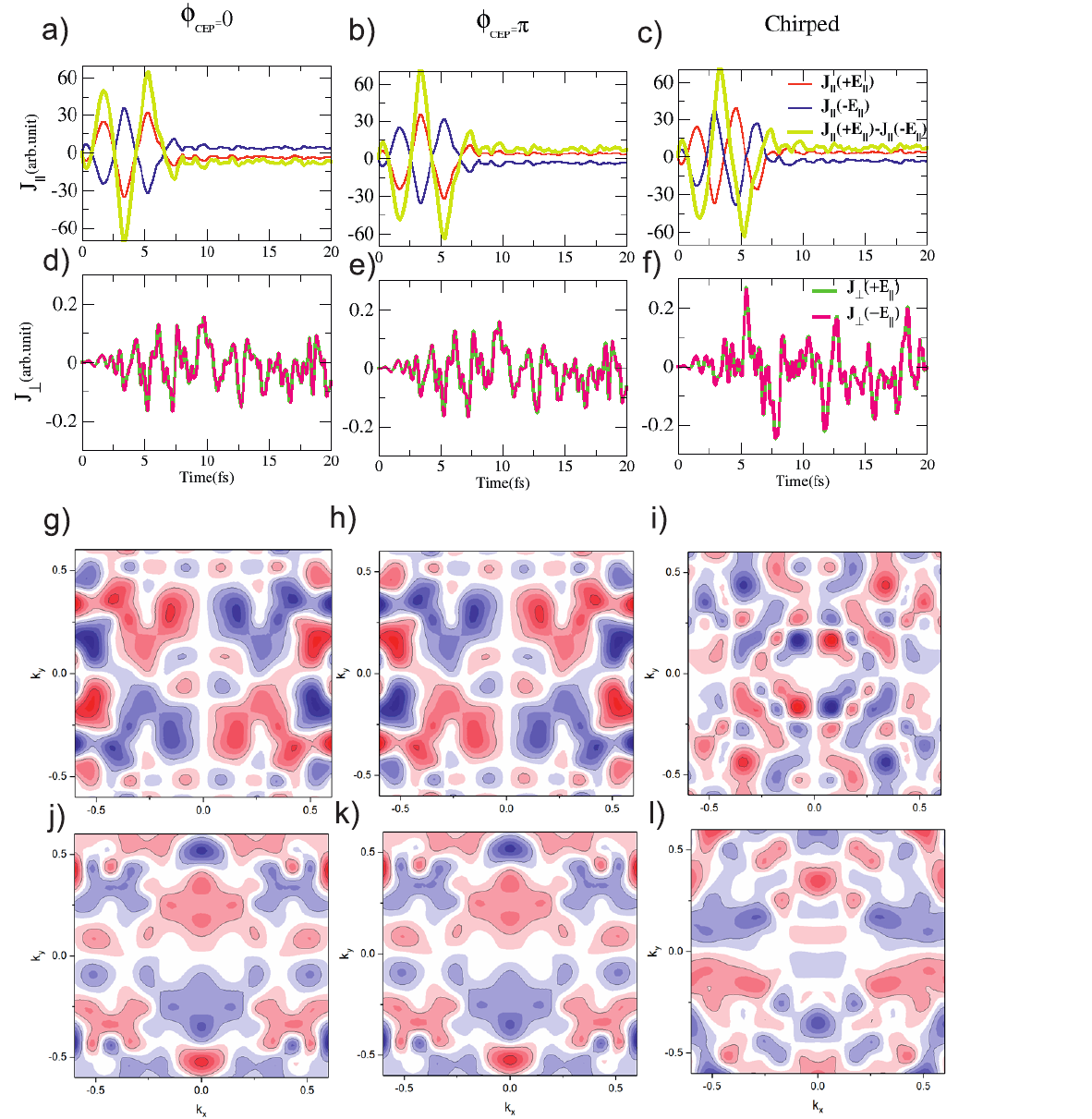}
    \caption{\textbf{Time and momentum-resolved profile of currents from chirp and CEP modulation}
    \textbf{(a)-(c)}Time-resolved profiles of longitudinal currents \( J_{\parallel} \)(induced current parallel to electric field)  and \textbf{(d)-(f)} transverse currents \( J_{\perp} \) (induced current perpendicular to electric field)  under opposite electric fields \( \mp E \) for unchirped (chirped) pulses and carrier-envelope phase $\phi_{CEP}$ values of 0 (\( \pi \)). The transverse current \( J_{\perp} \) does not change sign when the electric is reversed, indicating that it arises from a second-order nonlinear response. \textbf{(g)-(i)} show a comparison of the momentum-resolved distribution of \( j^{\Delta} \) at $\phi_{CEP}$ values of 0 and \( \pi \) and chirped pulse. The \( k \)-resolved \( j^{\Delta} \) exhibits odd parity of higher poles in the Berry curvature with respect to the mirror plane, when the current is reversed from $\phi_{CEP}$ = 0 to  \( \pi \) and chirped. \textbf{(j)-(l)} correspond to the \( k \)-resolved \( j^{\Sigma} \), as discussed in the text, which highlights the even-order poles of the Berry curvature. }
 \label{fig5}
\end{figure}

This compact formulation illustrates that the transverse current is comprised of a mixture of even and odd orders of interband Berry curvature multipoles. Additionally, the equation indicates the coupling of odd (even) orders of Berry curvature multipoles to odd (even) pseudoharmonics. Further, the Fourier transform of the current $\mathbf{J}_{\perp}(t)$ reveals that the current consists of both even and odd harmonics with a phase delay of $\pi/2$. The above equations \ref{even-sum}, \ref{even-odd-sum} and the associated arguments are particularly well-suited and valid for few-pulse dynamics induced by irradiation with sufficiently long pulses. Transitioning to short-pulse, subcycle dynamics likely breaks $\mathcal{T}$-symmetry, as the short pulses are generally asymmetric in time. Therefore, in our study, we restrict the use of equation (\ref{even-odd-sum}) to all phase-modulated cases. In light of this, $D^{(q)}$ for even (odd) $q$ must undergo transformation according to even (odd) parity by reversing the sign of $+E$ to $-E$, ensuring that $\mathbf{J}_{\perp}(t)$ does not change sign . This result aligns with the findings of Ye et al. \cite{ye2023control}, which demonstrate that the polarization direction of the Berry curvature is affected by the relative orientation of the electric field and the crystal axis. Notably, the Berry curvature direction can be reversed by changing the polarity of the DC field.
 Figures \cref{fig5}d-f depict the comparison between the transverse current under $\pm E \hat{\bm x}$  for $\phi_{CEP}$ of $0,\pi$, and  chirped pulse, respectively. Notably, the current for a $\phi_{CEP}$ of $\pi$ remains comparable to that for $\phi_{CEP}$ = 0, consistent with the summation described in Eq.~\ref{even-odd-sum}. Furthermore, the current under a chirped pulse, where the chirp phase further breaks time-reversal symmetry  $\mathcal{T}$, enhances the contribution from odd-order multipoles and harmonics. Extending the contribution beyond the Berry curvature dipole in enhanced nonlinear anomalous Hall effects due to $\mathcal{T}$ breaking is consistent with previous work \cite{holder2020consequences}.
 We then obtain the real-time sum and difference of transverse currents  as follows:

 \begin{equation}{\label{eq-2}}
\mathbf{J}^\Delta(t) = \frac{J_\perp^E - J_\perp^{-E}}{2} =
  0,
\end{equation}

and
\begin{equation}\label{eq-3}
\mathbf{J}^\Sigma(t) = \frac{J_\perp^E + J_\perp^{-E}}{2} = \sum_{\substack{p=\text{odd},\geq 1} }\sum_{q \geq p} 
\operatorname{Re} \left[e^{ip\Phi}\right] \frac{ \alpha_q E^q_0}{\Phi'^{q-1}} D^{(q)}+
 \sum_{\substack{p=\text{even}, \geq 2}} \sum_{q \geq p}
\operatorname{Im} \left[e^{ip\Phi}\right] \frac{E_0^q}{\Phi'^{q-1}} D^{(q)},
\end{equation}

   We can attribute these real time sum and difference currents to an effective $k$-resolved density current as $\mathbf{J}^{\Sigma( {\Delta})}(t)= \int_k dk \mathbf{j}^{\Sigma {(\Delta)}}(t,k) $.
 The \( k \)-resolved density current directly mirrors the distribution of Berry curvature in momentum space, with regions of high current intensity corresponding to areas of significant Berry curvature \cite{shin2019unraveling, luu2018measurement}. In Supplementary Materials Note 4, the analysis of \( k \)-resolved density currents $\mathbf{j}^{\Sigma {(\Delta)}}(t)$ under low-intensity linear and high-intensity nonlinear fields reveals that the X and Y valleys contribute predominantly in the weak field regime, while under strong fields, the entire Brillouin zone (BZ) becomes involved in the current.
 Figures.~\cref{fig5}g-l illustrate these momentum-resolved  density currents $\mathbf{j}^{(\Delta)}(t)$ and $\mathbf{j}^{\Sigma}(t)$ at the moment the laser is turned off.
 Comparing the $\mathbf{j}^{(\Delta)}(t)$ for unchirped($\phi_{CEP}$=0) with $\phi_{CEP}$=$\pi$ shows mirror-image opposites distribution coming from odd parity $D^{(q)}$ with odd $\textit{q}$ in Eq.\ref{even-odd-sum}. At low intensities, the odd parity term $D^{(1)}$ corresponds to $\Omega_{nm}$, while in the high-intensity regime, higher-order terms such as $\frac{\partial^2 \Omega}{\partial k^2}$ also become relevant. This holds true even in the chirped case (see \cref{fig5}l). The $k$-resolved distribution of the density $\mathbf{j}^{(\Delta)}(t)$ is somewhat different for $\phi_{CEP}$= $\pi$, as the chirped parameter further breaks time-reversal symmetry $\mathcal{T}$ and increases the frequency, making the occupation frequency-dependent. Nevertheless, odd parity terms $D^{(q)}$, such as the Berry curvature and its even derivatives, remain responsible for current and polarization reversal. As expected from the comparison of Figs.~\cref{fig5}j and \cref{fig5}k, the current $\mathbf{j}^{(\Sigma)}(t)$ reflects the even-parity $D^{(q)}$ terms, which remain unchanged under a reversal of the electric field sign.

\begin{figure}[t!]
    \centering
    \includegraphics[width=1\textwidth]{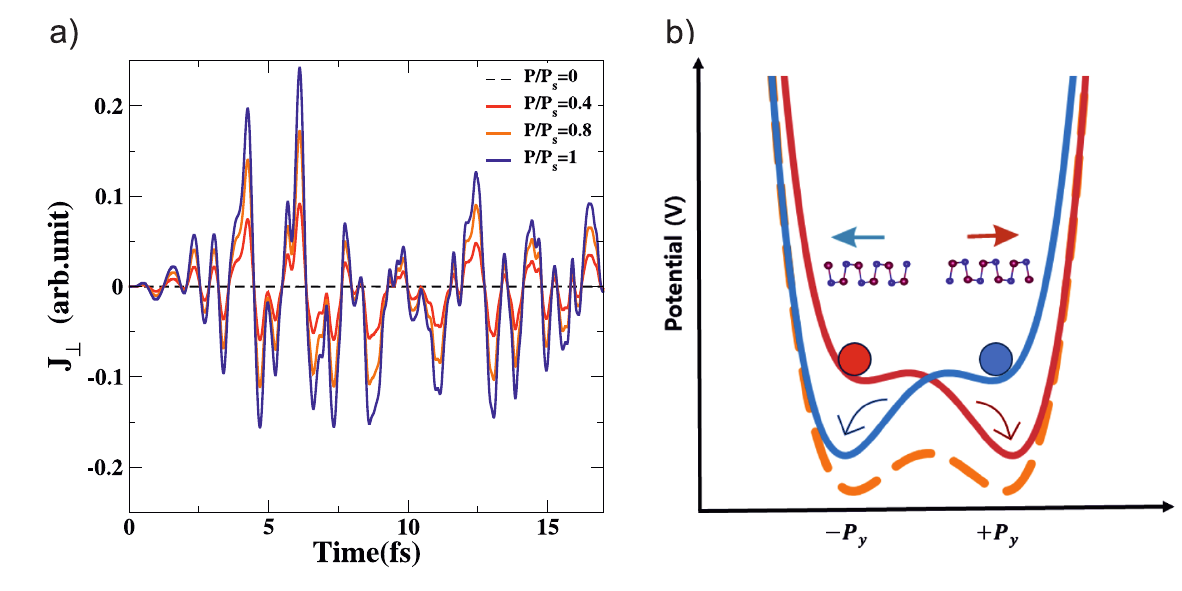}
    \caption{\textbf{Transverse Current and Chirped Pulse Effects on Ferroelectric PES}
    \textbf{(a)} The transverse current magnitude increases as higher-order contributions from the Berry curvature become more prominent, responding to the ferroelectric polarization strength. The polarization (P) scales from zero in the non-polar paraelectric phase to a maximum value ($P_s$) when the material reaches full ferroelectric polarization.  \textbf{(b)}The diagram shows how a chirped laser pulse changes the double-well potential energy surface (PES) of a ferroelectric material. The red curve represents the original (unchirped) PES, while the blue curve shows the modified (chirped) PES, creating an asymmetry. This asymmetry causes a polarization reversal from $P_{+y}$ to $P_{-y}$ after the laser is turned off. The orange curve represents the unperturbed PES.     
       }
 \label{fig-6}
\end{figure}

 Furthermore, to investigate the interplay between polarization and topology, we computed the transverse currents for varying magnitudes of ferroelectric polarization, ranging from 0 (the paraelectric phase) to 1 (the maximum ferroelectric phase), as illustrated in Fig.~\cref{fig-6}a. As shown, the enhancement of the transverse current is directly scaled and amplified with increasing polarization magnitude, driven by its correlation with the Berry curvature and its higher-order multipole components. These results emphasize the crucial role of ferroelectricity-induced topological effects in governing nonlinear optical responses, and highlight the significant contribution of higher-order Berry curvature terms to the observed current behavior.
The double-well potential maps the energy profile of varying polarization strengths, offering insights into polarization switching and stable polarization directions along the ferroelectric mode. Under laser exposure, this potential is modified, revealing dynamic behavior. For an unchirped pulse, symmetry between the wells is disrupted, stabilizing the \(+P_y\) well over the \(-P_y\) one. A chirped pulse reverses this preference, favoring the \(-P_y\) well As shown in Fig.~\cref{fig-6}b \cite{zhou2021terahertz, abalmasov2020ultrafast, hanakata2016polarization}. Both pulse types lower the barrier height between wells, facilitating polarization transitions. This dynamic switching between \(\pm P_y\) states, driven by the pulse type, offers a pathway for binary data storage, with stable polarization states representing 0 and 1 for future electronic devices.

\section{\small    MULTIBIT FERROELECTRICITY POLARIZATION CONTROL BY CHIRP AND HANDEDNESS}

\begin{figure}[t!]
    \centering
    \includegraphics[width=1\textwidth]{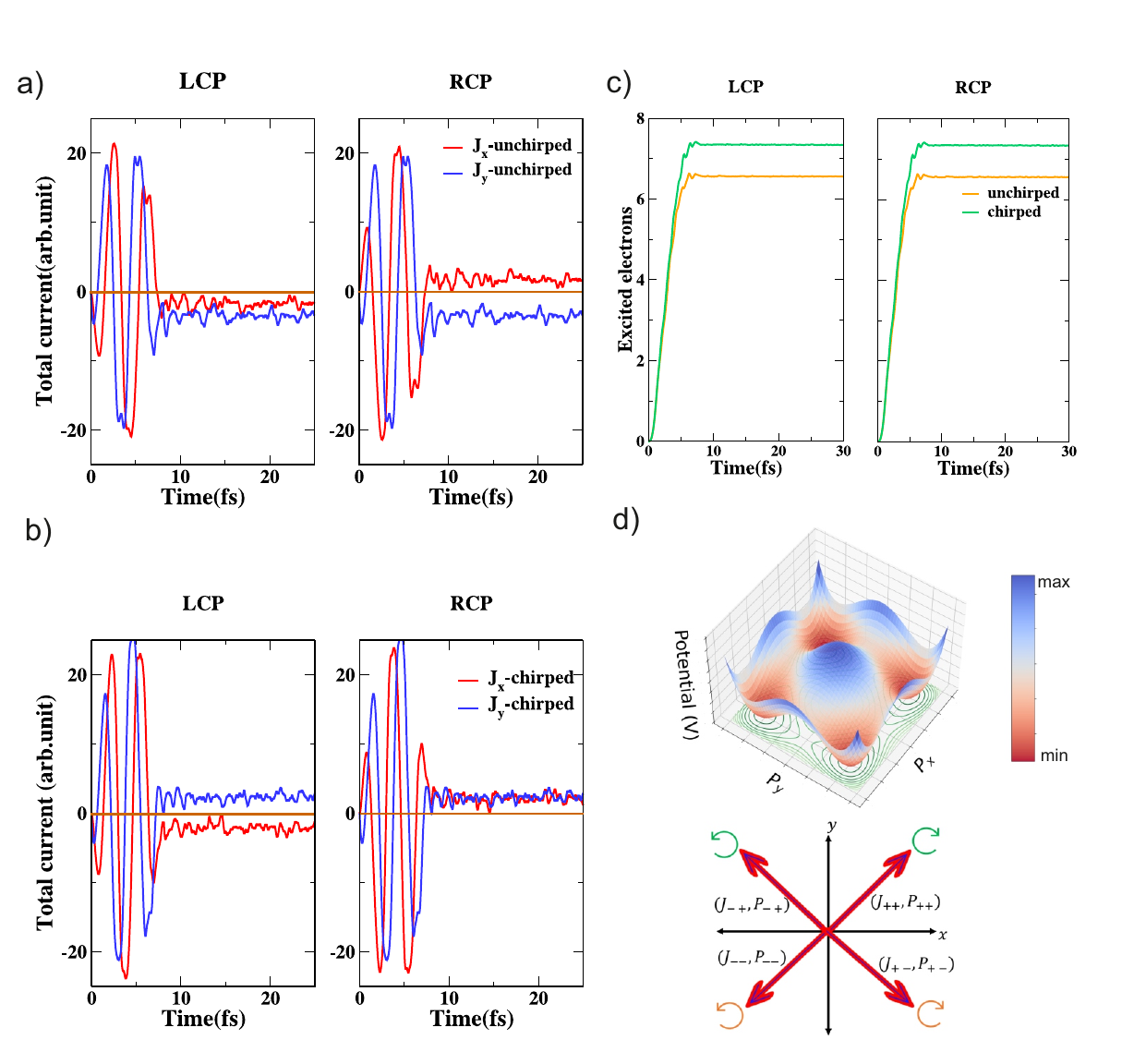}
    \caption{ \textbf{Chirp-Modulated Phase Effects in Circularly Polarized Light.}
    \textbf{(a)} and \textbf{(b)} show the effect of the chirp-modulated phase in circularly polarized light with right and left handedness, respectively, as compared to the unchirped case. \textbf{(c)} shows the electron population excited into the conduction band for both unchirped and chirped pulses of left and right circular polarization. \textbf{(d)}  demonstrates a \( \frac{\pi}{2} \)  rotation in the residual current and polarization for both unchirped (brown) and chirped (green) pulses in left- and right-circularly polarized (LCP and RCP) light. Four distinct ferroelectric states arise from combinations of light handedness and the chirp parameter, each defined within a four-well potential landscape.
}
 \label{fig4}
\end{figure}

We now compare the current induced by the unchirped and chirped circularly polarized pulses, as illustrated in Fig.~\cref{fig4}a and b. As expected, changing the handedness from left-handed circular polarization (LCP) to right-handed circular polarization (RCP) reverses the direction of the residual current components. Specifically, \( J_x \) switches from negative to positive, while \( J_y \) consistently remains negative. However, for chirped circularly polarized light (CPL), reversing the polarization from LCP to RCP results in \( J_y \) always being positive, with \( J_x \) changing sign from negative to positive. This comparison highlights that, while the total residual current undergoes a \( \frac{\pi}{2} \) rotation when switching between unchirped LCP and RCP, introducing a chirp phase to CPL induces an additional \( \frac{\pi}{2} \) rotation relative to the unchirped case. This combined effect of light helicity and chirp phase enables the realization of four distinct ferroelectricity states corresponding to four well potential picture, as schematically illustrated in Fig.~\cref{fig4}d, compared to linearly polarized light  that supports only two states, corresponding to a \( \pi \) rotation in the current and polarization direction.

A previous study \cite{wang2019ferroicity} demonstrated a \( \pi \)-phase shift in the current direction under both linearly and circularly polarized light for shift and injection currents. In contrast, when coupled with ferroelastic transitions along the \( x \)- or \( y \)-axis of a monolayer, both shift and injection currents undergo a \( 90^\circ \) rotation under the same polarization conditions. In this work, we show that chirping the pulse breaks time-reversal symmetry $\mathcal{T}$, enabling the residual current to dictate the direction of the injection current. Consequently, similar to the findings of the earlier study, linearly polarized chirped light results in a \( 180^\circ \) reversal of both the shift current and the residual injection current. However, chirped circularly polarized light induces a \( 90^\circ \) rotation of the total residual currents, accompanied by a \( 90^\circ \) reversal in polarization.

\section{\small DISCUSSION}

In our study, we investigated the phase modulation of strong-field, short-duration pulses in the nonlinear photocurrent responses of 2D ferroelectric monolayer SnTe. By applying a time-varying chirp and constant carrier-envelope phase ($\phi_{CEP}$) offset, we observed significant imbalance of population within the Brillouin zone (BZ) as a function of the chirp rate.  Notably, we found a directional reversal of the residual current upon removal of the laser field. This current reversal was observed for both chirped and constant CEP phase-modulated light, irrespective to its linear or circular polarization. Importantly, this reversal of the injection current was accompanied by a concomitant rotation in the polarization direction. 

 Time-resolved analysis of injection currents, as a function of the incident angle of irradiation, indicates that the chirp parameter can effectively control the direction of polarization by modifying the regional occupation weight across the BZ. Additionally, the \( k \)-resolved second-order transverse current scales with the Berry curvature even and odd-order poles and pseudo-harmonics in strong-field regimes. As the external electric field dynamically modulates the ferroelectric polarization, these higher-order Berry curvature moments contribute to complex nonlinear optical phenomena such as higher harmonic generation and nonlinear Hall effects. Moreover, exploiting the circularly polarized driver, both through the manipulation of chirp parameter or handedness of the circular driver, we can effectively adjust the time-reversal symmetry $\mathcal{T}$ of the system, leading to non-equivalent interactions with the electronic states. This results in ferroelectricity multi-states with distinct current and polarization characteristics.
  This interaction enhances the sensitivity of polarization dynamics to external stimuli, particularly  to sub-laser-cycle dynamics ,potentially improving the material's nonvolatile memory capabilities through the coupling of electronic states and ferroelectric order. We anticipate that the observed reversal and rotation of current and polarization, driven by $\phi_{chirp}$ or $\phi_{CEP}$ modulation, could be applicable to group-IV monochalcogenide monolayers due to their similar electronic and atomic structures. These findings suggest potential extensions beyond monolayer SnTe, offering new strategies for phase-controlled nonlinear optics in other 2D materials and broadening the scope of ultrafast optoelectronic applications. Unlike previous studies that focused primarily on static field effects, our work demonstrates dynamic manipulation of photocurrents and polarization, providing a more versatile approach to ferroelectric control.

\section{\small METHOD}

Our calculations utilize time-dependent density functional theory (TD-DFT) to model the time-evolution of the wave function and electronic current within the adiabatic local-density approximation (ALDA), as implemented in the real-space, real-time framework via the OCTOPUS code \cite{octopus-1, octopus-2, octopus-3}. We examine electron dynamics under the influence of both linearly and circularly polarized laser pulses, incorporating either a time-dependent chirp parameter or a constant carrier-envelope phase $\phi_{CEP}$. The electrons dynamics are captured by solving the time-dependent Kohn-Sham equations in the velocity gauge.

\begin{equation}\label{eq-1}
i\hbar \frac{\partial \phi_n (r,t)}{\partial t} =  \left\{\frac{1}{2m}\left[-i\hbar \bm{\nabla} +\frac{e}{c} \bm{A}(t)\right]^2 +V_{ion}(r,t)+\int dr' \frac{e^2}{|r-r'|}\rho(r',t) + V_{xc}(r,t) \right\}\phi_n(r,t),
\end{equation}

where  $\bm{A}(t)$  represents the time-dependent profile of the vector potential, which is written as

\begin{equation}
  \bm{A}=A_0 f(t) \cos(\omega t+\Phi) \hat{\bm{e}}
\end{equation}

Here, \( f(t) \) represents the envelope function, \( \omega \) denotes the driver frequency, and \( \hat{e} \) specifies the polarization vector, which can be either linearly or circularly polarized. The phase \( \Phi \) of the electric waveform is tailored to support two pulse-shaping methods: a dynamic phase characterized by a linear chirp, \( \Phi_{chirp} = b t^2 \) with \( b \) ranging from \( [0.01, 0.1] \, \text{fs}^{-2} \), and a constant carrier-envelope phase $\phi_{CEP}$, which introduces a phase offset between the carrier wave and the pulse envelope. Both $\Phi_{chirp}$ and $\phi_{CEP}$  laser modulation break time-reversal symmetry, thereby altering the electric field profile within the pulse.

We set the strong intensity  $I=10^{13} W/cm^2$ for incident pulse with the duration of two cycles, ensuring the sub-cycle dynamics. The driving frequency was set to $\omega=$1.55 eV , which surpasses the band gap and corresponds to one of the prominent peaks in the shift current spectra.\cite{jin2024peculiar}. Ionic motion and phonon excitation are fully frozen in the present work under the assumption that the pulse durations are on the order of femtoseconds, significantly shorter than typical lattice dynamics response times. We minimally tested the effects of phonon, by exciting \( A_u \) and \( E_u \) infrared-active modes. Though the phonons contribute to the amplifications of polarization dynamics,  (see Supplementary Materials Note 3), they do not fundamentally alter the current reversal behavior with which we are mainly concerned in the present work. It is important to note that the lattice's ionic component lacks sufficient time to respond to these extremely short pulses.

The real-space grid is sampled with a spacing of 0.35 Bohr, and the \( k \)-point mesh for the 2D Brillouin zone is uniformly set to \( 24 \times 24 \times 1 \) based on convergence tests. The momentum-resolved current density within the unit cell is calculated as:

\begin{equation}
  \mathbf{j}(t,k)=\sum_{n} \frac{1}{\Omega} Re \left[ \int d\mathbf{r} \phi^*_{nk}(r,t) \hat{\mathbf{v}} \phi_{nk}(r,t)  \right]
\end{equation}

The average electric current density over the entire Brillouin zone is then given by:

\begin{equation}
  \mathbf{J}(t)=\sum_{k} \sum_{n} \frac{1}{\Omega} Re \left[ \int d\mathbf{r} \phi^*_{nk}(r,t) \hat{\mathbf{v}} \phi_{nk}(r,t)  \right]
\end{equation}

where \( \hat{\mathbf{v}}=\frac{1}{m_e} \left( -i\hbar \bm{\nabla} +\frac{e}{c} \mathbf{A} \right) \) is the velocity operator, and \( \Omega \) represents the unit cell volume.

The number of excited electrons \( N_{ex} \) in the conduction band is obtained as:

\begin{equation}
N_{ex}(t)=  \sum_{k} \sum_{n=n_{occ}+1}^{\infty} \sum_{m=1}^{n_{occ}} \left|\left\langle \phi_{m,k}(t>0)|\phi_{n,k}(t=0)\right\rangle\right|^2
\label{n-ex-eq}
\end{equation}

where $n$ is the band index and $k$ is the Bloch number, respectively.

\section{\small DATA AVAILABILITY}
The data that support the findings of this study are available from the corresponding
author upon reasonable request.

\newpage
\bibliography{ref}
\section{\small ACKNOWLEDGEMENTS}
This research was supported by Brain Pool program funded by the Ministry of Science and ICT through the National Research Foundation of Korea(RS-2023-00283963)
\section{\small COMPETING INTERESTS}
The authors declare no competing interests.


\section*{Supplementary material}

# Multistate Control of Nonlinear Photocurrents in Optoferroelectrics via phase manipulation of light field

Ali Kazempour,[†] Esmaeil Taghizadeh Sisakht,[‡] Mahmut Sait Okyay,[¶] Xiao Jiang,[‡] Shunsuke Sato,[§] and Noejung Park[*,‡]

[†]Department of physics, Payame Noor University, PO BOX 119395-3697, Tehran, Iran

[‡]Department of Physics, Ulsan National Institute of Science and Technology, Ulsan, 689-798, Korea

[¶]Materials Science and Engineering Program, Department of Chemical and Environmental Engineering, University of California-Riverside, Riverside, CA 92521, USA

[§]Center for Computational Sciences, University of Tsukuba,Tsukuba 305-8577 Japan

E-mail: noejung@unist.ac.kr

## Note 1:injection Current reversal under chirpe phase modulation in group IV-monochalcogenides

To investigate the generalization of current reversal using the chirp parameter as a control variable, we calculate a similar structure for the GeSe monolayer, which has a band gap of 1.12 eV, approximately twice that of SnTe's gap (0.7 eV). Under conditions similar to those of SnTe, with a chirp rate parameter of $b = 0.01\,\text{fs}^{-2}$, the residual current induced by the chirped pulse is reversed relative to the unchirped pulse, as shown in Fig. 1S. However, the

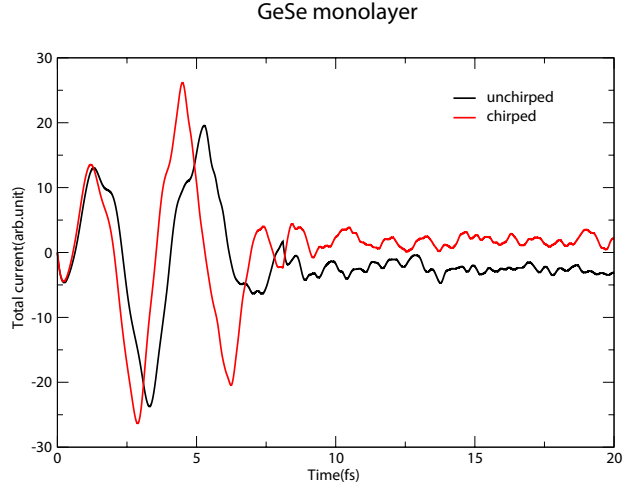

Figure S1: The time-dependent current induced in a GeSe monolayer, which has a similar puckered structure but a larger band gap compared to the SnTe monolayer, shows a similar reversal of the remaining injected current when the pulse is chirped.

shift current and residual injection current are both lower than in SnTe. Notably, as the band gap increases, the magnitude of the induced second-order injection current decreases.

## Note 2: The effect of different negative and positive chirp rate

To gain a quantitative understanding of how different rates of chirping influence the residual injection current, we compare the total current profiles of the SnTe monolayer under negative and positive chirp rates of 0.005, 0.01, 0.03, 0.05, and 0.1 $\text{fs}^{-2}$, as well as the number of excited electrons injected into the conduction band. As illustrated in Fig. S2, the same values of $\pm b$ rates result in different magnitudes and directions of the residual current. Therefore, chirp rates $b$ that provide opposite signs of the vector potential before the laser ends can reverse the current. It is noteworthy that the same value of chirp rate, but with opposite

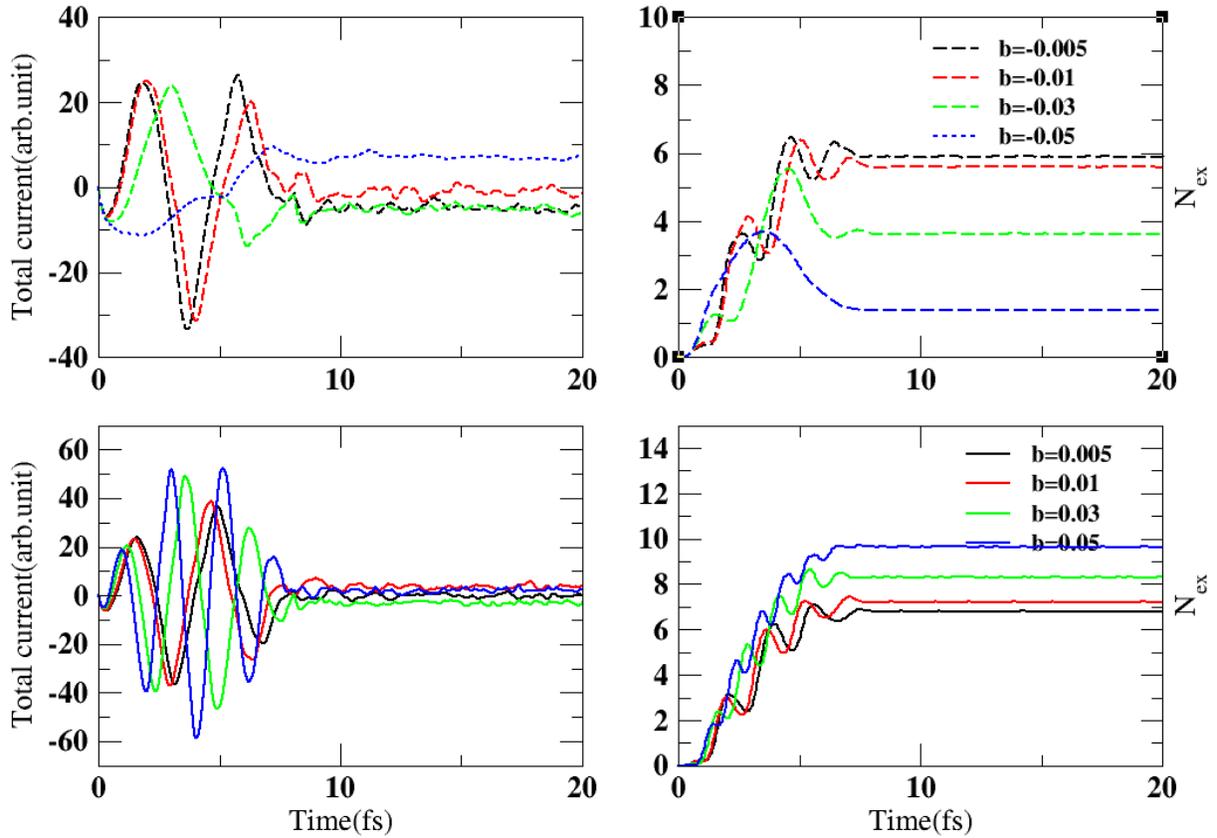

Figure S2: . The time evolution of the current and the number of excited electrons in the conduction band are shown for both negative (top) and positive (bottom) chirp rates. Even with negative chirp rates, there is still a significant residual current, which corresponds to the conduction band being occupied by a single electron (shown by the blue dashed line).

signs, results in very different numbers of excited electrons. The current reversal is attributed to changes in the k-space population distribution, which affect the balance between intraband and interband processes.

# Note 3: Intensity dependence and ionic relaxation and nonlinear phononics

To assess the second-order character of the residual current, we present the intensity of the driver on the residual current. Fig. S4.a shows the intensity dependence of the remnant

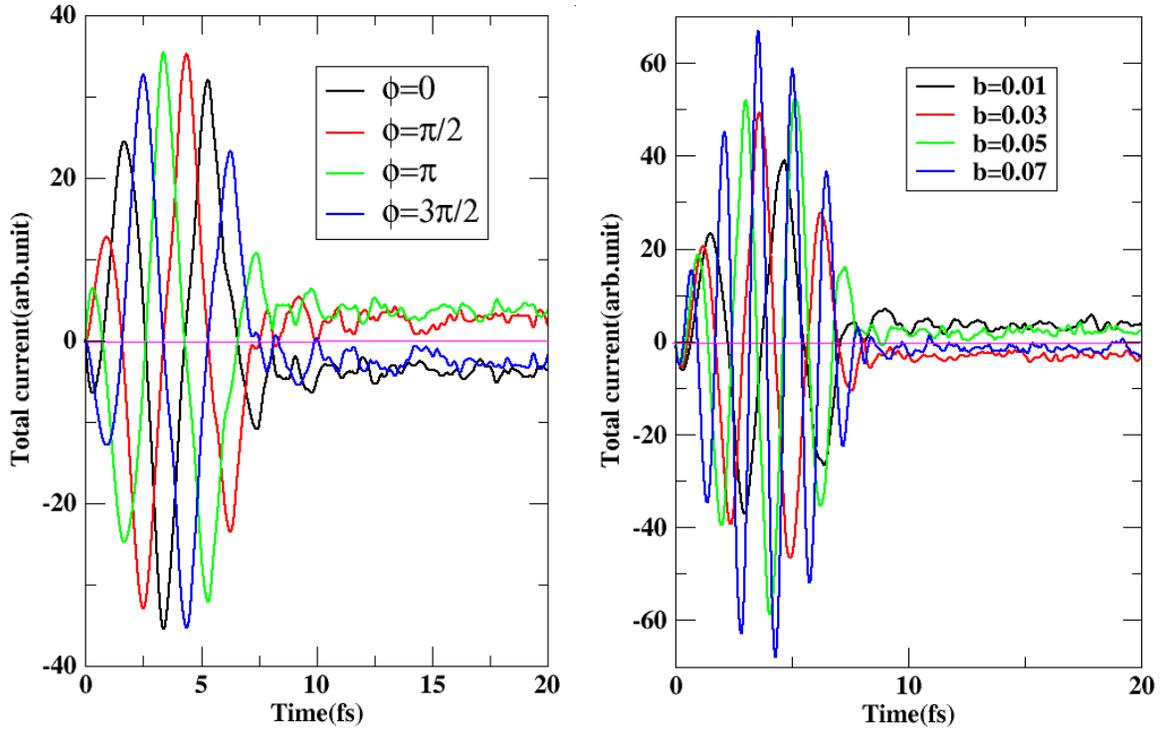

Figure S3: The current alternates direction (switching from negative to positive and back) due to phase modulation of the laser pulse. This behavior depends on the different chirp rates and fixed carrier-envelope phase (CEP) values.

current for three different intensities. It is clear that the residual current reverses from a positive value with a low-intensity oscillatory character to a negative finite value at high intensity. This behavior indicates that light-induced electron dynamics at strong field intensities introduce highly nonlinear excitation channels, such as tunneling paths, leading to a substantial population imbalance within the entire momentum space. In other words, at low intensity, the k-space population is low, and the electron population remains mostly near the band edges, while at high intensity, the k-space population becomes dense and asymmetric, resulting in a significant population of states far from the band edges.

Fig. s3.b further presents a comparison of the chirped light-induced residual current with the unchirped one when electron-phonon scattering is allowed. Electrons can scatter off

4

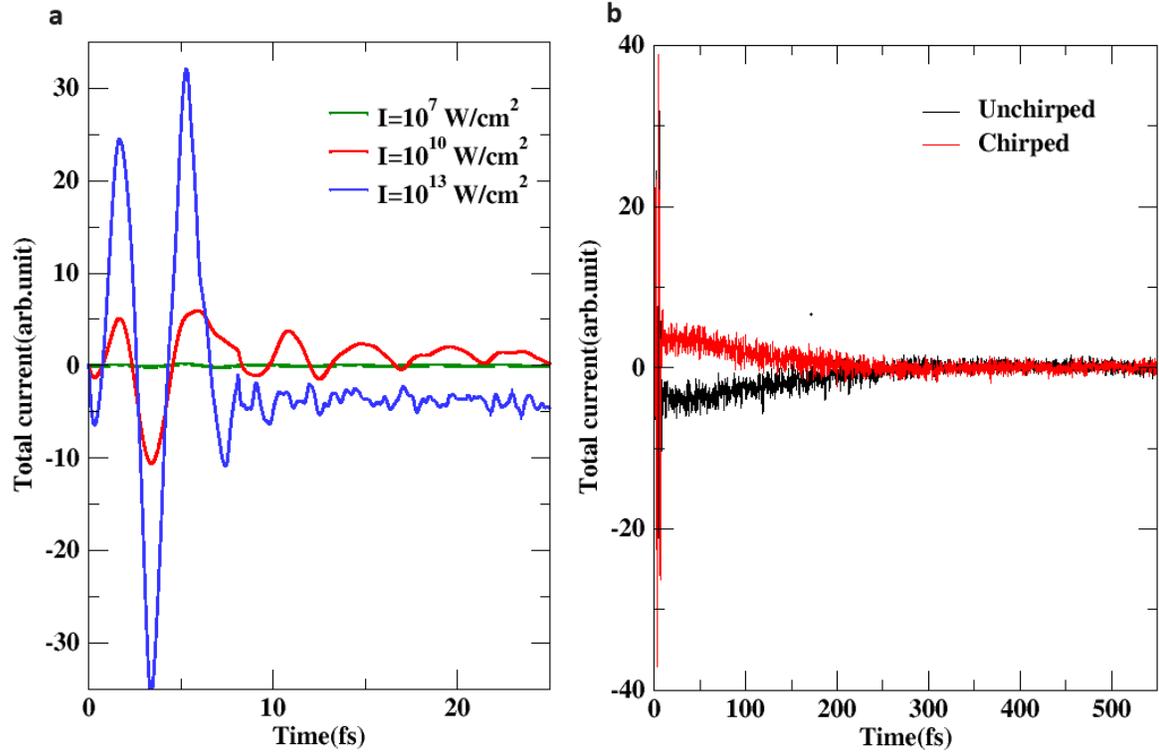

Figure S4: . Current behavior depends on the field intensity, shown for weak, medium, and strong fields (left). At high field intensities, the remaining current saturates at a negative value, unlike the unsaturated positive current seen at low field intensities. The time-dependent long lasting residual current around 250 fs is also shown with and without ionic relaxation included in the calculations for unchirped (chirped) pulses (right).

vibrating ions, leading to changes in their momentum and energy, which affect the current. The residual carrier relaxation decays around 250 fs for both unchirped and chirped lasers at the high intensity limit. While chirped pulses can lead to temporal stretching of the excitation and impact the initial carrier distribution, the residual relaxation (after initial scattering events) tends to converge for both pulse types because the decay is driven by electron-phonon interactions, which dominate in the high intensity regime.

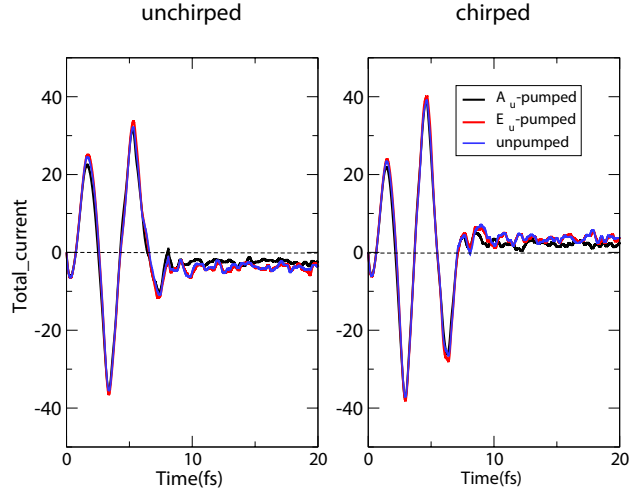

Figure S5: . Au and Eu pumped unchirped and unchirped currents

# Note 4:low intensity k-resolved current and BC poles derivation in broken TRS

Fig. S6 shows the comparison of the $J^\Delta$ distribution of residual currents for three incident intensities of chirped laser light. It is worth noting that at sufficiently high intensity, all regions in the entire Brillouin zone contribute to the light-induced current, while at low intensity, the valleys are the most significant contributors to the current. Moreover, the appearance of different parts of the Brillouin zone contributing to the induced current indicates the formation of higher-order Berry curvature poles, as discussed in.[1] As mentioned in,[1] when time-reversal symmetry (TRS) is broken, the Berry curvature no longer needs to be an odd function of crystal momentum, which leads to an asymmetric distribution of Berry curvature and enhances the Berry curvature dipole, inducing a transverse current that results in significant nonlinear Hall effects.

Starting from Eq. (2) of the main article, in the presence of an electric field given by

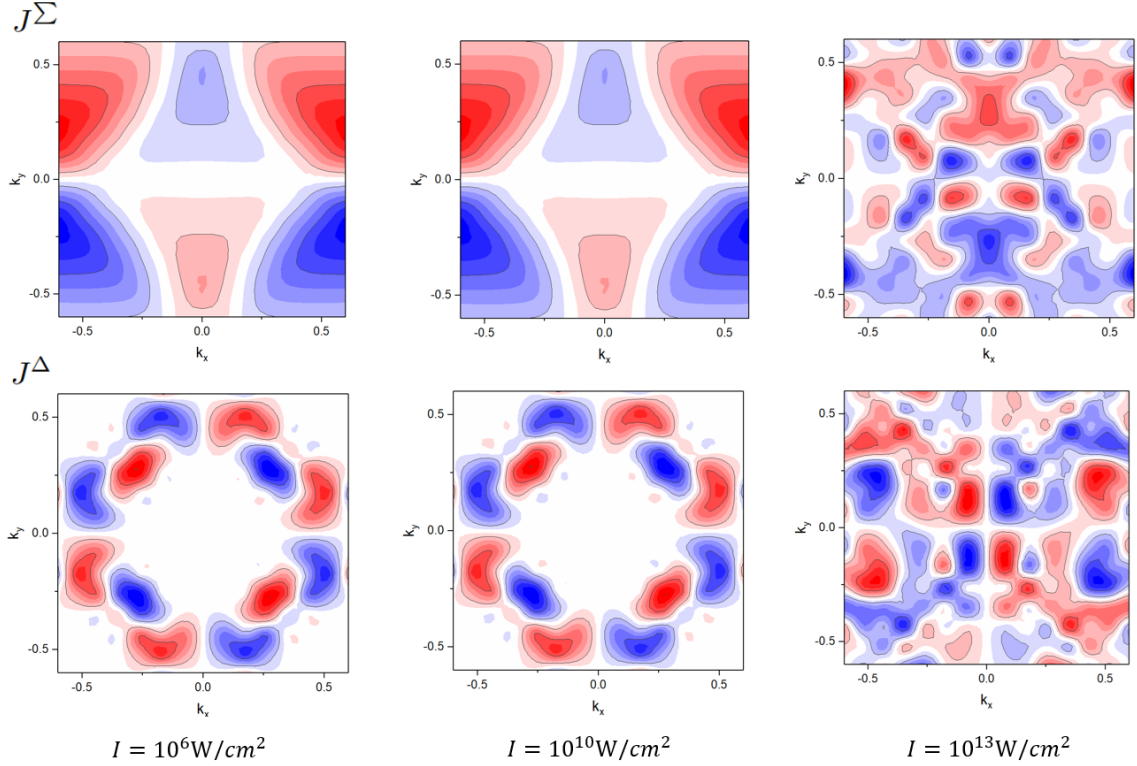

Figure S6: . From left to right: The momentum-resolved distribution of $J^\Delta$ and $J^\Sigma$ for weak, moderate, and strong fields is shown. In the strong field limit, higher-order Berry curvature effects are generated.

$\boldsymbol{E}(t) = E_0 \operatorname{Re}\left[e^{i\Phi(t)}\right]\hat{\boldsymbol{x}}$, and considering the Bloch oscillation $\Delta k = \frac{-eE_0}{\hbar \Phi'(t)} \operatorname{Im}\left[e^{i\Phi(t)}\right]$, which arises from the Peierls substitution, our goal is to expand $\Omega_{nm}$ in terms of powers of $\Delta k$:

$$\Omega_{nm}(k + \Delta k_x) = \Omega_{nm}(k) + \frac{\partial \Omega_{nm}(k)}{\partial k_x}\Delta k_x + \frac{1}{2}\frac{\partial^2 \Omega_{nm}(k)}{\partial k_x^2}\Delta k_x^2 + \\ \frac{1}{6}\frac{\partial^3 \Omega_{nm}(k)}{\partial k_x^3}\Delta k_x^3 + \frac{1}{24}\frac{\partial^4 \Omega_{nm}(k)}{\partial k_x^4}\Delta k_x^4 + \frac{1}{120}\frac{\partial^5 \Omega_{nm}(k)}{\partial k_x^5}\Delta k_x^5 + \cdots \tag{S1}$$

Considering that odd (even) power of $\Delta \boldsymbol{k}$ multiplying the electric field $cos(\Phi)sin^p(\Phi)$ is written as series of $\sum\limits_{p=odd} sin(n\Phi)$ for even integers of n and $\sum\limits_{p=even} cos(n'\Phi)$ for odd integers $n'$. Combining all terms together we will find the following equation for $\mathcal{T}$-conserved case:

$$\mathbf{J}_\perp(t) = -Im[e^{2i\Phi}]\{\alpha\frac{E_0^2}{\Phi'}\boldsymbol{D}_{nm}^{(2)}+\beta\frac{E_0^4}{\Phi'^3}\boldsymbol{D}_{nm}^{(4)}+\cdots\}-Im[e^{4i\Phi}]\{\beta'\frac{E_0^4}{\Phi'^3}\boldsymbol{D}_{nm}^{(4)}+\gamma'\frac{E_0^6}{\Phi'^5}\boldsymbol{D}_{nm}^{(6)}+\cdots\}+\cdots.$$
(S2)

where $\Omega_{nm}(\mathbf{k}+\Delta k\hat{x}) + \Omega_{nm}(-\mathbf{k}+\Delta k\hat{x}) = 2\frac{\partial\Omega_{nm}}{\partial k_x}\Delta k_x$

For $\mathcal{T}$-broken case the transverse current is written as

$$\mathbf{J}_\perp(t) = Re[e^{i\Phi}]\{\alpha E_0 \boldsymbol{D}_{nm}^{(1)} + \beta\frac{E_0^3}{\Phi'^2}\boldsymbol{D}_{nm}^{(3)} + \cdots\} + Re[e^{3i\Phi}]\{\beta'\frac{E_0^3}{\Phi'^2}\boldsymbol{D}_{nm}^{(3)} + \gamma'\frac{E_0^5}{\Phi'^4}\boldsymbol{D}_{nm}^{(5)} + \cdots\}$$
$$-Im[e^{2i\Phi}]\{\alpha\frac{E_0^2}{\Phi'}\boldsymbol{D}_{nm}^{(2)} + \beta\frac{E_0^4}{\Phi'^3}\boldsymbol{D}_{nm}^{(4)} + \cdots\} - Im[e^{4i\Phi}]\{\beta'\frac{E_0^4}{\Phi'^3}\boldsymbol{D}_{nm}^{(4)} + \gamma'\frac{E_0^6}{\Phi'^5}\boldsymbol{D}_{nm}^{(6)} + \cdots\}+\cdots.$$
(S3)

Where its compact forms of two above equations are written in the main text.


\end{document}